\documentclass{ws-procs9x6}
\def\bc{\begin{center}}
\def\ec{\end{center}}
\def\bea{\begin{eqnarray}}
\def\eea{\end{eqnarray}}

\def\gappeq{\mathrel{\rlap {\raise.5ex\hbox{$>$}} {\lower.5ex\hbox{$\sim$}}}}
\def\lappeq{\mathrel{\rlap{\raise.5ex\hbox{$<$}} {\lower.5ex\hbox{$\sim$}}}}
\def\MeV{{\rm MeV}}\def\GeV{{\rm GeV}}
\def\swsqeffl{\sin^2{\theta_{eff}}}
\def\dalhad{\Delta\alpha^{(5)}_{had}(m_Z)}

\begin{document}

\title{New Physics and the LHC\footnote{\uppercase{W}e recognize that this work has been partly supported by  the \uppercase{I}talian \uppercase{M}inistero dell'\uppercase{U}niversita' e della \uppercase{R}icerca \uppercase{S}cientifica, under the \uppercase{COFIN} program for 2007-08.
}}

\author{Guido Altarelli}

\address{Dipartimento di Fisica `E.~Amaldi', Universit\`a di Roma Tre
and INFN, Sezione di Roma Tre, I-00146 Rome, Italy and \\ CERN, Department of Physics, Theory Unit, 
 CH--1211 Geneva 23, Switzerland}

\maketitle

\abstracts{
In these lectures I start by briefly reviewing the status of the electroweak theory, in the Standard Model and beyond. I then discuss the motivation and the possible avenues for new physics, on the brink of the LHC start.}

\begin{flushright}
{RM3-TH/08-9}\\   
{CERN-PH-TH/2008-085}\\
\end{flushright}

\section{The programme of LHC physics}

The first collisions at the LHC are expected in '08 and the physics run at $14$ TeV will start soon after. The particle physics community eagerly waits for the answers that one expects from the LHC  to a number of big questions. The main physics issues at the LHC, addressed by the ATLAS and CMS collaborations, will be: 1) the experimental clarification of the Higgs sector of the electroweak (EW) theory, 2) the search for new physics at the weak scale that, on conceptual grounds, one predicts should be in the LHC discovery range, and 3) the identification of the particle(s) that make the dark matter in the Universe. In addition the LHCb detector will be devoted to the study of precision B physics, with the aim of going deeper in the knowledge of the Cabibbo-Kobayashi-Maskawa (CKM) matrix and of CP violation. The LHC will also devote a number of runs to accelerate heavy ions and the ALICE collaboration will study their collisions for an experimental exploration of the QCD phase diagram.

\section{The Higgs problem}

The Higgs problem is really central in particle physics today. On the one hand, the experimental verification of the Standard Model (SM) cannot be considered complete until the physics of the  Higgs sector is not established by experiment. On the other hand, the Higgs is directly related to most of the major open problems of particle physics, like the flavour problem or the hierarchy problem, the latter strongly suggesting the need for new physics near the weak scale (which could possibly clarify the dark matter identity). It is clear that the fact that some sort of Higgs mechanism is at work has already been established. The W or the Z with longitudinal polarization that we observe are not present in an unbroken gauge theory (massless spin-1 particles, like the photon, are transversely polarized). The longitudinal degree of freedom for the W or the Z is borrowed from the Higgs sector and is an evidence for it. Also, the couplings of quarks and leptons to
the weak gauge bosons W$^{\pm}$ and Z are indeed precisely those
prescribed by the gauge symmetry.  To a lesser
accuracy the triple gauge vertices $\gamma$WW and ZWW have also
been found in agreement with the specific predictions of the
$SU(2)\bigotimes U(1)$ gauge theory. This means that it has been
verified that the gauge symmetry is unbroken in the vertices of the
theory: all currents and charges are indeed symmetric. Yet there is obvious
evidence that the symmetry is instead badly broken in the
masses. Not only the W and the Z have large masses, but the large splitting of, for example,  the t-b doublet shows that even a global weak SU(2) is not at all respected by the fermion spectrum. Symmetric coupling and completely non symmetric spectrum are a clear signal of spontaneous
symmetry breaking which, in a gauge theory, is implemented via the Higgs mechanism. The big remaining questions are about
the nature and the properties of the Higgs particle(s).

The present experimental information on the Higgs sector, mainly obtained from LEP as described in section 4, 
is surprisingly limited. It can be summarized in a few lines, as follows. First, the relation $M_W^2=M_Z^2\cos^2{\theta_W}$, modified by small, computable
radiative corrections, has been
experimentally proven. This relation means that the effective Higgs
(be it fundamental or composite) is indeed a weak isospin doublet.
The Higgs particle has not been found but, in the SM, its mass can well
be larger than the present direct lower limit $m_H \gappeq 114.4$~GeV (at $95\%$ c.l.)
obtained from  searches at LEP-2.  As we shall see, the radiative corrections
computed in the SM when compared to the data on precision electroweak
tests lead to a clear indication for a light Higgs, not too far from
the present lower bound. The experimental upper limit on $m_H$, obtained from fitting the data in the SM, depends on the value of the top quark mass $m_t$ (the one-loop radiative corrections are quadratic in $m_t$ and logarithmic in $m_H$).  The CDF and D0 combined value after Run II is at present\cite{ewwg} $m_t= 172.6\pm1.4~GeV$ (it went  down with respect to the value $m_t= 178\pm4.3~GeV$ from Run I and also the experimental error is now sizably reduced). As a consequence the present limit on $m_H$ is more stringent: $m_H < 190~GeV$ (at $95\%$ c.l., after including the information from the 114.4 GeV direct bound). On the Higgs the LHC will address the following questions : do the Higgs particles actually exist? How many: one doublet, several doublets, additional singlets? SM Higgs or SUSY Higgses? Fundamental or composite (of fermions, of WW...)? Pseudo-Goldstone boson of an enlarged symmetry? A manifestation of large extra dimensions (5th component of a gauge boson, an effect of orbifolding or of boundary conditions...)? Or some combination of the above or something so far unthought of?

\section{Theoretical bounds on the SM Higgs}

The LHC has been designed to solve the Higgs puzzle. In the SM lower and upper limits on the Higgs mass can be derived from theoretical considerations. It is well known\cite{zzi},\cite{zzii},\cite{aaiiii} that in the SM with only one Higgs doublet a lower limit on
$m_H$ can be derived from the requirement of vacuum stability (or, in milder form, from a moderate instability, compatible with the lifetime of the Universe \cite{isi}). The limit is a function of $m_t$ and of the energy scale
$\Lambda$ where the SM model breaks down and new physics appears. The Higgs mass enters because it fixes the initial value of the quartic Higgs coupling $\lambda$ for its running up to the large scale $\Lambda$. Similarly an upper bound on $m_H$ (with mild dependence
on $m_t$) is obtained\cite{eeiiii} from the requirement that in $\lambda$, up to the scale $\Lambda$, no Landau pole appears, or in more explicit terms, that the perturbative description of the theory remains valid.  The upper limit on the Higgs mass in the SM is clearly important for assessing the chances of success of the LHC as an accelerator designed to solve the Higgs problem. Even if 
$\Lambda$ is as small as ~a few TeV the limit is $m_H < 600-800~$GeV and becomes $m_H < 180~$GeV for
$\Lambda \sim M_{Pl}$. We now briefly recall the derivation of these limits.
	
The possible instability of the Higgs potential $V[\phi]$ is generated by the quantum loop corrections to the
classical expression of $V[\phi]$. At large $\phi$ the derivative $V'[\phi]$ could become negative and the potential
would become unbound from below. The one-loop corrections to $V[\phi]$ in the SM are well known and change the dominant
term at large $\phi$ according to $\lambda \phi^4 \rightarrow (\lambda +
\gamma~{\rm log}~\phi^2/\Lambda^2)\phi^4$. The one-loop approximation is not enough in this case, because it fails
at large enough $\phi$, when
$\gamma~{\rm log}~\phi^2/\Lambda^2$ becomes of order 1. The renormalization group improved version of the corrected
potential leads to the replacement $\lambda\phi^4 \rightarrow
\lambda(\Lambda)\phi'^4(\Lambda)$ where $\lambda(\Lambda)$ is the running coupling and
$\phi'(\mu) =\phi {\rm exp}\int^t \gamma(t')dt'$, with $\gamma(t)$ being an anomalous dimension function and $t = {\rm
log}\Lambda/v$ ($v$ is the vacuum expectation value
$v = (2\sqrt 2 G_F)^{-1/2}$). As a result, the positivity condition for the potential amounts to the requirement that
the running coupling $\lambda(\Lambda)$ never becomes negative. A more precise calculation, which also takes into
account the quadratic term in the potential, confirms that the requirements of positive
$\lambda(\Lambda)$ leads to the correct bound down to scales $\Lambda$ as low as $\sim$~1~TeV. The running of
$\lambda(\Lambda)$ at one loop is given by: 
\begin{equation}
\frac{d\lambda}{dt} = \frac{3}{4\pi^2} [ \lambda^2 + 3\lambda h^2_t - 9h^4_t + {\rm small~gauge~and~Yukawa~terms }]~,
\label{131}
\end{equation} with the normalization such that at $t=0, \lambda = \lambda_0 = m^2_H/2v^2$ and, for the top Yukawa coupling,
$h_t^0 = m_t/v$. We see that, for $m_H$ small and $m_t$ fixed at its measured value,
$\lambda$ decreases with $t$ and can become negative.  If one requires that
$\lambda$ remains positive up to $\Lambda = 10^{15}$--$10^{19}$~GeV, then the resulting bound on $m_H$ in the SM with
only one Higgs doublet is given by\cite{aaiiii}:
\begin{equation} m_H(\rm{GeV}) > 132 + 2.1 \left[ m_t - 172.6 \right] - 4.5~\frac{\alpha_s(m_Z) - 0.118}{0.006}~.
\label{25h}
\end{equation}
Note that this limit is evaded in models with more Higgs doublets. In this case the limit applies to some average mass but the lightest Higgs particle can well be below, as it is the case in the minimal SUSY extension of the SM (MSSM).

The upper limit on
the Higgs mass in the SM is clearly important for assessing the chances of success of the LHC as an accelerator designed
to solve the Higgs problem. The upper limit\cite{eeiiii} arises from the requirement that the Landau pole associated with the non
asymptotically free behaviour of the $\lambda \phi^4$ theory does not occur below the scale $\Lambda$. The initial value of
$\lambda$ at the weak scale increases with $m_H$ and the derivative is positive at large $\lambda$ (because of the positive $\lambda^2$ term in eq.(\ref{131}) - the $\lambda \varphi^4$ theory is not asymptotically free - which overwhelms the negative top-Yukawa term). Thus if $m_H$ is too large the point where $\lambda$, computed from the perturbative beta function, becomes infinite (the Landau pole) occurs at too low an energy.  Of course in the vicinity of the Landau pole the 2-loop evaluation of the beta function is not reliable. Indeed the limit indicates the frontier of the domain where the theory is well described by the perturbative  expansion. Thus the quantitative evaluation of the limit is only indicative, although it has been to some extent supported by simulations of the Higgs sector of the EW theory on the lattice. For the upper limit on $m_H$ one finds\cite{eeiiii}
$m_H\lappeq 180~GeV$ for $\Lambda\sim M_{GUT}-M_{Pl}$ and $m_H\lappeq 0.5-0.8~TeV$ for $\Lambda\sim
1~TeV$. Actually, for
$m_t \sim$ 172~GeV, only a small range of values for $m_H$ is allowed, $130 < m_H <\sim 200$~GeV, if the SM holds up
to $\Lambda \sim M_{GUT}$ or $M_{Pl}$.  An additional argument indicating that the solution of the Higgs problem cannot be too far away is the fact that, in the absence of a Higgs particle or of an alternative mechanism, violations of unitarity appear in scattering amplitudes involving longitudinal gauge bosons (those most directly related to the Higgs sector) at energies in the few TeV range\cite{unit}. In conclusion, it is very unlikely that the solution of the Higgs problem can be missed at the LHC which has a good sensitivity up to $m_H\sim 1~{\rm TeV}$.

\section{Precision tests of the standard electroweak theory}

The most precise tests of the electroweak theory apply to the QED sector. The anomalous magnetic moments of the electron and of the muon are among the most precise measurements in the whole of physics. Recently there have been new precise measurements of $a_e$ and $a_\mu$ for the electron\cite{ae1} and the muon\cite{amu} ($a = (g-2)/2$). On the theory side, the QED part has been computed analytically for $i=1,2,3$, while for $i=4$ there is a numerical calculation (see, for example, ref.\cite{kino}). Some terms for $i=5$ have also been estimated for the muon case. The weak contribution is from $W$ or $Z$ exchange. The hadronic contribution is from vacuum polarization insertions and from light by light scattering diagrams.  For the electron case the weak contribution is essentially negligible and the hadronic term does not introduce an important uncertainty.  As a result the $a_e$ measurement can be used to obtain the most precise determination of the fine structure constant\cite{ae2}. In the muon case the experimental precision is less by about 3 orders of magnitude, but the sensitivity to new physics effects is typically increased by a factor $(m_\mu/m_e)^2 \sim 4^.10^4$. The dominant theoretical ambiguities arise from the hadronic terms in vacuum polarization and in light by light scattering. If the vacuum polarization terms are evaluated from the $e^+e^-$ data a discrepancy of $\sim 3 \sigma$ is obtained (the $\tau$ data would indicate better agreement, but the connection to $a_\mu$ is less direct and recent new data have added solidity to the $e^+e^-$ route)\cite{amu2}. Finally, we note that, given the great accuracy of the $a_\mu$ measurement and the estimated size of the new physics contributions, for example from SUSY, it is not unreasonable that a first signal of new physics would appear in this quantity.

\begin{figure}[ht]
\centerline{\epsfxsize=3.1in\epsfbox{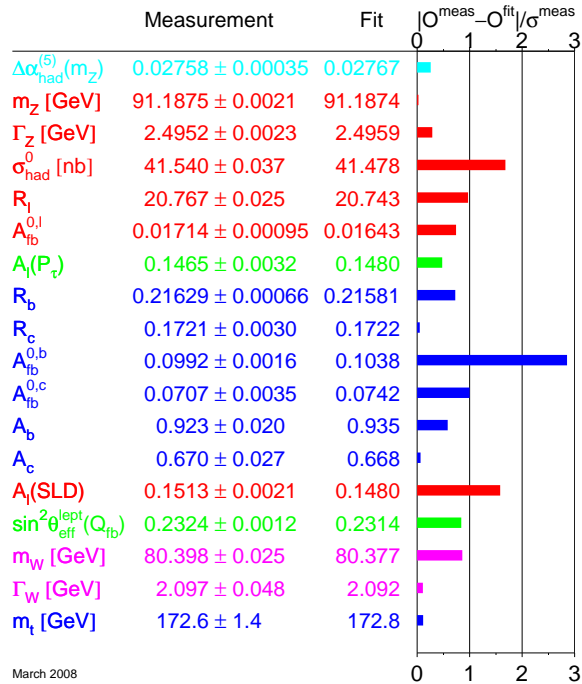}}   
\caption{Precision tests of the Standard EW theory from LEP, SLC and the TeVatron (March'08). \label{pull}}
\end{figure}

The results of the electroweak precision tests as well as of the searches
for the Higgs boson and for new particles performed at LEP and SLC are now available in  final form\cite{ewwg}.  Taken together
with the measurements of $m_t$, $m_W$ and the searches for new physics at the Tevatron, and with some other data from low
energy experiments, they form a very stringent set of precise constraints to be compared with the SM or with
any of its conceivable extensions\cite{AG}. 
All high energy precision tests of the SM are summarized in fig.~\ref{pull}\cite{ewwg}.
For the analysis of electroweak data in the SM one starts from the
input parameters: as in any renormalizable theory masses and couplings
have to be specified from outside. One can trade one parameter for
another and this freedom is used to select the best measured ones as
input parameters. Some of them, $\alpha$, $G_F$ and
$m_Z$, are very precisely known, some other ones, $m_{f_{light}}$,
$m_t$ and $\alpha_s(m_Z)$ are far less well determined while $m_H$ is
largely unknown.  Among the light fermions, the quark masses are badly known, but
fortunately, for the calculation of radiative corrections, they can be
replaced by $\alpha(m_Z)$, the value of the QED running coupling at
the Z mass scale. The value of the hadronic contribution to the
running, $\dalhad$, reported in Fig.~\ref{pull}, is obtained
through dispersion relations from the data on $e^+e^-\rightarrow \rm{hadrons}$ at
low centre-of-mass energies~\cite{ewwg}. From the input parameters
one computes the radiative corrections to a sufficient precision to
match the experimental accuracy. Then one compares the theoretical
predictions with the data for the numerous observables which have been
measured, checks the consistency of the theory and derives constraints
on $m_t$, $\alpha_s(m_Z)$ and $m_H$.

The computed radiative corrections include the complete set of
one-loop diagrams, plus some selected large subsets of two-loop
diagrams and some sequences of resummed large terms of all orders
(large logarithms and Dyson resummations). In particular large
logarithms, e.g., terms of the form $(\alpha/\pi ~{\rm
ln}~(m_Z/m_{f_\ell}))^n$ where $f_{\ell}$ is a light fermion, are
resummed by well-known and consolidated techniques based on the
renormalisation group. For example, large logarithms dominate the
running of $\alpha$ from $m_e$, the electron mass, up to $m_Z$, which
is a $6\%$ effect, much larger than the few per mil contributions of
purely weak loops.  Also, large logs from initial state radiation
dramatically distort the line shape of the Z resonance observed at
LEP-1 and SLC and have been accurately taken into account in the
measurement of the Z mass and total width.

Among the one loop EW radiative corrections a remarkable class of
contributions are those terms that increase quadratically with the top
mass.  The large sensitivity of radiative corrections to $m_t$ arises
from the existence of these terms. The quadratic dependence on $m_t$
(and possibly on other widely broken isospin multiplets from new
physics) arises because, in spontaneously broken gauge theories, heavy
loops do not decouple. On the contrary, in QED or QCD, the running of
$\alpha$ and $\alpha_s$ at a scale $Q$ is not affected by heavy quarks
with mass $M \gg Q$. According to an intuitive decoupling
theorem~\cite{AC}, diagrams with heavy virtual particles of mass
$M$ can be ignored for $Q \ll M$ provided that the couplings do not
grow with $M$ and that the theory with no heavy particles is still
renormalizable. In the spontaneously broken EW gauge theories both
requirements are violated. First, one important difference with
respect to unbroken gauge theories is in the longitudinal modes of
weak gauge bosons. These modes are generated by the Higgs mechanism,
and their couplings grow with masses (as is also the case for the
physical Higgs couplings). Second, the theory without the top quark is
no more renormalizable because the gauge symmetry is broken if the b
quark is left with no partner (while its couplings show that the weak
isospin is 1/2). Because of non decoupling precision tests of the
electroweak theory may be sensitive to new physics even if the new
particles are too heavy for their direct production.

While radiative corrections are quite sensitive to the top mass, they
are unfortunately much less dependent on the Higgs mass. If they were
sufficiently sensitive, by now we would precisely know the mass of the
SM Higgs. In fact, the dependence of one loop diagrams on $m_H$ is only
logarithmic: $\sim G_F m_W^2\log(m_H^2/m_W^2)$. Quadratic terms $\sim
G_F^2 m_H^2$ only appear at two loops and are too small to be
important. The difference with the top case is that $m_t^2-m_b^2$ is a
direct breaking of the gauge symmetry that already affects the
relevant one loop diagrams, while the Higgs couplings to gauge bosons
are "custodial-SU(2)" symmetric in lowest order.

\begin{figure}[ht]
\centerline{\epsfxsize=3.6in\epsfbox{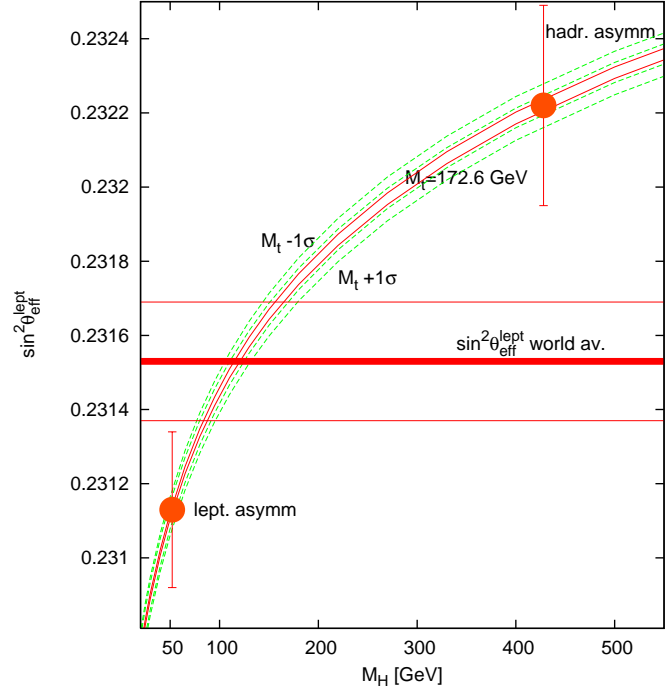}}   
\caption{The data for $\sin^2\theta_{\rm eff}^{\rm lept}$ are
plotted vs $m_H$. For presentation purposes the measured points are
shown each at the $m_H$ value that would ideally correspond to it
given the central value of $m_t$. \label{s2mH}}
\end{figure}

The various asymmetries determine the effective electroweak mixing
angle for leptons with highest sensitivity.   The weighted
average of all results, including small correlations, is:
\begin{eqnarray}
\swsqeffl & = & 0.23153\pm0.00016 \,.
\label{eq:sin2teff}
\end{eqnarray}
Note, however, that this average has a $\chi^2$ of 11.8 for 5 degrees
of freedom, corresponding to a probability of 3.7\%. The $\chi^2$ is
pushed up by the two most precise measurements of $\swsqeffl$, namely
those derived from the measurements of $A_l$ by SLD, dominated by the
left-right asymmetry $A_{LR}$, and of the forward-backward asymmetry
measured in $b \bar b$ production at LEP, $A^b_{FB}$, which differ by about
3.2 $\sigma$'s.   In
general, there appears to be a discrepancy between $\swsqeffl$
measured from leptonic asymmetries ($(\sin^2\theta_{\rm eff})_l$) and
from hadronic asymmetries ($(\sin^2\theta_{\rm eff})_h$), as seen from Figure~\ref{s2mH}. In fact, the result from $A_{LR}$ is in good
agreement with the leptonic asymmetries measured at LEP, while all
hadronic asymmetries, though their errors are large, are better
compatible with the result of $A^b_{FB}$.
This very unfortunate fact makes the interpretation of precision tests less sharp and some perplexity remains: is it an experimental error or a signal of some new physics?

The situation is shown in Figure~\ref{s2mH}~\cite{Gam}.
The values of $(\sin^2\theta_{\rm eff})_l$, $(\sin^2\theta_{\rm
eff})_h$ and their formal combination are shown each at the $m_H$
value that would correspond to it given the central value of $m_t$.
Of course, the value for $m_H$ indicated by each $\swsqeffl$ has an
horizontal ambiguity determined by the measurement error and the width
of the $\pm1\sigma$ band for $m_t$.  Even taking this spread into
account it is clear that the implications on $m_H$ are sizably
different.  

One might imagine that some new physics effect could be
hidden in the $\mathrm{Z b \bar b}$ vertex.  Like for the top quark
mass there could be other non decoupling effects from new heavy states
or a mixing of the b quark with some other heavy quark.  However, it
is well known that this discrepancy is not easily explained in terms
of some new physics effect in the $\mathrm{Z b \bar b}$ vertex. In
fact, $A^b_{FB}$ is the product of lepton- and b-asymmetry factors:
$A^b_{FB}=(3/4)A_e A_b$.  The sensitivity of $A^b_{FB}$ to $A_b$ is
limited, because the $A_e$ factor is small, so that a rather large
change of the b-quark couplings with respect to the SM is needed in
order to reproduce the measured discrepancy (precisely a $\sim 30\%$
change in the right-handed coupling $g^b_R$, an effect too large to be a loop
effect but which could be produced at the tree level, e.g., by mixing
of the b quark with a new heavy vectorlike quark\cite{CTW} or of the $Z$ with an heavier $Z'$\cite{zepr}).  But
this effect is not confirmed by the direct measurement
of $A_b$ performed at SLD using the left-right polarized b asymmetry, which agrees with the precision within the moderate precision of this result. Also, no deviation is manifest in the accurate measurement of $R_b \propto
g_{\mathrm{Rb}}^2+g_{\mathrm{Lb}}^2$ (but there $g^b_R$ is not dominant).   Thus, even introducing an
ad hoc mixing the overall fit of $A^b_{FB}$, $A_b$ and $R_b$ is not terribly good, but we cannot
exclude the possibility of new physics completely.  Alternatively, the observed
discrepancy could be due to a large statistical fluctuation or an
unknown experimental problem. In any case the effective ambiguity in the measured value of
$\swsqeffl$ is actually larger than the nominal error, reported in
Eq.~\ref{eq:sin2teff}, obtained from averaging all the existing
determinations.

We now discuss fitting the data in the SM. One can think of different
types of fit, depending on which experimental results are included or
which answers one wants to obtain. For example\cite{ewwg}, in
Table~\ref{tab:fit:result} we present in column~1 a fit of all Z pole
data plus $m_W$, $\Gamma_W$ (this is interesting as it shows the value
of $m_t$ obtained indirectly from radiative corrections, to be
compared with the value of $m_t$ measured in production experiments),
in column~2 a fit of all Z pole data plus $m_t$ (here it is $m_W$
which is indirectly determined), and, finally, in column~3 a fit of
all the data listed in Fig. 1 (which is the most
relevant fit for constraining $m_H$).  From the fit in column~1 of
Table~\ref{tab:fit:result} we see that the extracted value of $m_t$ is
in good agreement with the direct measurement (see the value reported in
Fig. 1).  Similarly we see that the direct
determination of $m_W$ reported in Fig. 1 is still a bit larger with respect to the value from the fit in
column~2 (although the direct value of $m_W$ went down recently).  We have seen that quantum corrections depend only
logarithmically on $m_H$.  In spite of this small sensitivity, the
measurements are precise enough that one still obtains a quantitative
indication of the Higgs mass range in the SM. From the fit in column~3 we obtain:
$\log_{10}{m_H(\GeV)}=1.94\pm 0.16$ (or $m_H=87^{+36}_{-27}~\GeV$). We see that the central value of $m_H$ from the fit is below the lower limit on the SM Higgs mass from direct searches $m_H\gappeq 114~\GeV$, but within $1\sigma$ from this bound. If we had reasons to remove the result on $A^b_{FB}$  from the fit, the fitted value of $m_H$ would move down to something like: $m_H=55^{+30}_{-20}~\GeV$, further away from the lower limit.
\begin{table}[tb]
\begin{center}
\renewcommand{\arraystretch}{1.3}
\begin{tabular}{|l||c|c|c|}
\hline 
Fit       & 1 & 2 & 3 \\
\hline
\hline
Measurements      &$m_W$         &$m_t$            &$m_t,~m_W$\\
\hline
\hline
$m_t~(\GeV)$      &$178.7^{+12}_{-9}$&$172.6\pm1.4$    &$172.8\pm1.4$\\
$m_H~(\GeV)$      &$143^{+236}_{-80}$    &$111^{+56}_{-39}$&$87^{+36}_{-27}$\\
$\log~[m_H(\GeV)]$&$2.16\pm{+0.39}$&$2.05\pm0.18$    &$1.94\pm0.16$ \\
$\alpha_s(m_Z)$   &$0.1190\pm0.0028$     &$0.1190\pm0.0027$&$0.1185\pm0.0026$\\
\hline
$m_W~(\MeV)$      &$80385 \pm 21$    &$80363 \pm 20$   &$80377 \pm 15$  \\
\hline
\end{tabular}
\caption[]{ Standard Model fits of electroweak data. All fits use the
Z pole results and $\dalhad$ as listed in Fig.~\ref{pull}. In
addition, the measurements listed on top of each column are included as
well. The fitted W mass is also shown\cite{ewwg} (the directly measured value is
$m_W=80398 \pm 25 \MeV $).}
\label{tab:fit:result}
\end{center}
\end{table} 

\begin{figure}[ht]
\centerline{\epsfxsize=4.1in\epsfbox{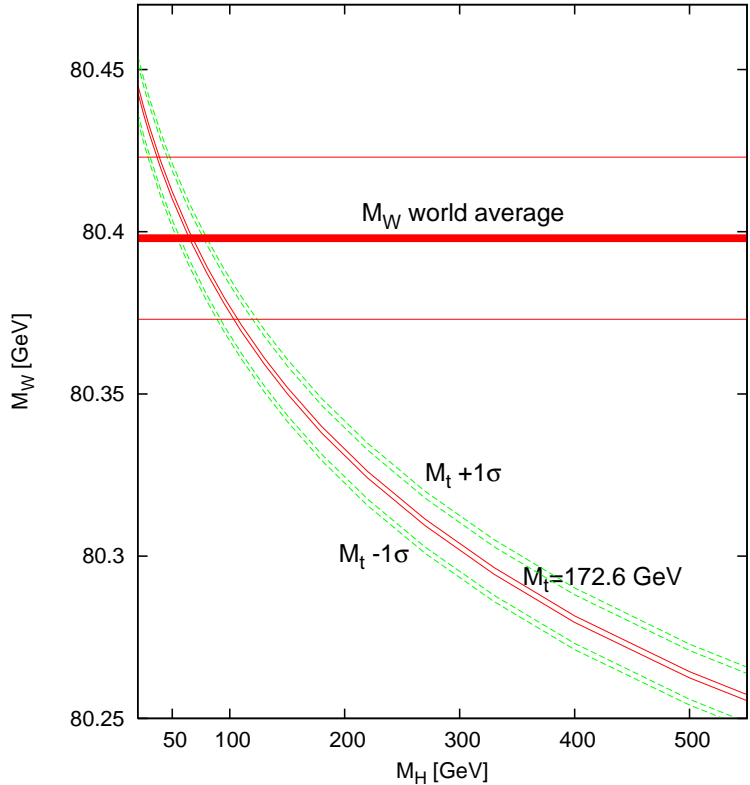}}   
\caption{The world average for $m_W$ is
plotted vs $m_H$. \label{wmH}}
\end{figure}

We have already observed that the experimental value of $m_W$ (with
good agreement between LEP and the Tevatron) is a bit high compared to
the SM prediction (see Figure~\ref{wmH},\cite{Gam}). The value of $m_H$
indicated by $m_W$ is on the low side, just in the same interval as
for $\sin^2\theta_{\rm eff}^{\rm lept}$ measured from leptonic
asymmetries.  The recent decrease of the experimental value of $m_t$
maintains the tension between the experimental
values of $m_W$ and $\sin^2\theta_{\rm eff}^{\rm lept}$ measured from
leptonic asymmetries on the one side and the lower limit on $m_H$ from
direct searches on the other side~\cite{cha},\cite {acggr}.  
 
With all these words of caution in mind it remains true that on the whole the SM performs
rather well, so that it is fair to say that no clear indication for
new physics emerges from the data. Actually the result of precision tests on the Higgs mass is particularly remarkable. The value of
$\log_{10}{[m_H(\GeV)]}$ is, within errors, inside the small window between
$\sim 2$ and $\sim 3$ which is allowed, on the one side, by the direct
search limit ($m_H\gappeq 114.4~\GeV$ from LEP-2~\cite{ewwg}),
and, on the other side, by the theoretical upper limit on the Higgs
mass in the minimal SM \cite{eeiiii}, $m_H\lappeq 600-800~\GeV$.

Thus the whole picture of a perturbative theory with a fundamental
Higgs is well supported by the data on radiative corrections. It is
important that there is a clear indication for a particularly light
Higgs: at $95\%$ c.l. $m_H\lappeq 190~\GeV$.  This is quite
encouraging for the ongoing search for the Higgs particle.  More in
general, if the Higgs couplings are removed from the Lagrangian the
resulting theory is non renormalizable. A cutoff $\Lambda$ must be
introduced. In the quantum corrections $\log{m_H}$ is then replaced by
$\log{\Lambda}$ plus a constant. The precise determination of the
associated finite terms would be lost (that is, the value of the mass
in the denominator in the argument of the logarithm).  A heavy Higgs
would need some conspiracy or some dynamical reason\cite{bar06}: the finite terms, different in the new theory from those of the SM, should accidentally or dynamically compensate
for the heavy Higgs in a few key parameters of the radiative
corrections (mainly $\epsilon_1$ and $\epsilon_3$, see, for example,
\cite{eps}).  Alternatively, additional new physics, for example in
the form of effective contact terms added to the minimal SM
lagrangian, should do the compensation, which again needs
some sort of conspiracy or some special dynamics, although this possibility is not so unlikely to be apriori discarded.

\section{The physics of flavour}

\begin{figure}[ht]
\centerline{\epsfxsize=4.1in\epsfbox{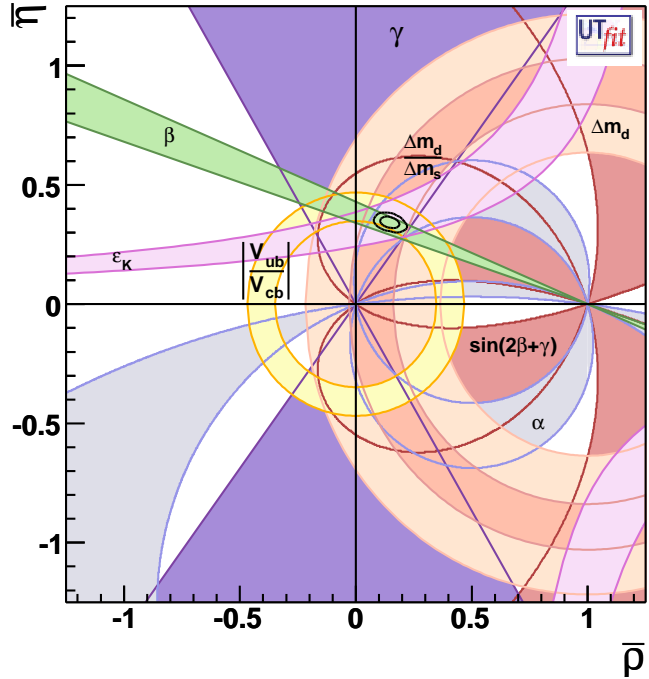}}   
\caption{Constraints in the $\bar \rho,\bar \eta$ plane including  the most recent $\alpha$, $\gamma$ and $\Delta M_s$ inputs in the global CKM fit. \label{CKM}}
\end{figure}

In the last decade great progress in different areas of flavour physics has been achieved. In the quark sector, the amazing results of a generation of frontier experiments, obtained at B factories and at accelerators, have become available\cite{fle}. QCD has been playing a crucial role in the interpretation of experiments by a combination of effective theory methods (heavy quark effective theory, NRQCD, SCET), lattice simulations and perturbative calculations. A great achievement obtained by many theorists over the last years is the calculation at NNLO of the branching ratio for $B\rightarrow X_s \gamma$ with B a beauty meson\cite{mis}. The effect of the photon energy cut, $E_\gamma > E_0$, necessary in practice, has been evaluated at NNLO\cite{bec}. The central value of the theoretical prediction is now slightly below the data: for $B[B\rightarrow X_s\gamma, E_0=1.6~GeV](10^{-4})$ the experimental value is 3.55(26)\cite{hfag} and the theoretical value is 3.15(23)\cite{mis} or 2.98(26)\cite{bec}, which to me is good agreement.
The hope of the B-decay experiments was to detect departures from the CKM picture of mixing and of CP violation as  signals of new physics. Finally, in quantitative terms, all measurements are in agreement with the CKM description of mixing and CP violation as shown in Fig. \ref{CKM}\cite{CKMfitter}. The recent measurement of $\Delta m_s$ by CDF and D0, in fair agreement with the SM expectation, has closed another door for new physics. But in some channels, especially those which occur through penguin loops, it is well possible that substantial deviations could be hidden (possible hints are reported in $B\rightarrow K\pi$ decays\cite{nat} and in $b\rightarrow s$ transitions\cite{ciu}). But certainly the amazing performance of the SM in flavour changing  and/or CP violating transitions in K and B decays poses very strong constraints on all proposed models of new physics\cite{isid}. 

In the leptonic sector the study of neutrino oscillations has led to the discovery that at least two neutrinos are not massless and to the determination of the mixing matrix\cite{alfe}. Neutrinos are not all massless but their masses are very small (at most a fraction of $eV$). Probably masses are small because $\nu ^\prime$s are Majorana fermions, and, by the see-saw mechanism, their masses are inversely proportional to the large scale $M$ where lepton number ($L$) non conservation occurs (as expected in GUT's). Indeed the value of $M\sim m_{\nu R}$ from experiment is compatible with being close to $M_{GUT} \sim 10^{14}-10^{15}GeV$, so that neutrino masses fit well in the GUT picture and actually support it. The interpretation of neutrinos as Majorana particles enhances the importance of experiments aimed at the detection of neutrinoless double beta decay and a huge effort in this direction is underway\cite{avi}.  It was realized that decays of heavy $\nu_R$ with CP and L non conservation can produce a B-L asymmetry. The range of neutrino masses indicated by neutrino phenomenology turns out to be perfectly compatible with the idea of baryogenesis via leptogenesis\cite{buch}. This elegant model for baryogenesis has by now replaced the idea of baryogenesis near the weak scale, which has been strongly disfavoured by LEP. 

It is remarkable that we now know the neutrino mixing matrix with good accuracy. Two mixing angles are large and one is small. The atmospheric angle $\theta_{23}$ is large, actually compatible with maximal but not necessarily so: at $3\sigma$\cite{mal}: $0.34 \leq \sin^2{\theta_{23}}\leq 0.68$ with central value around  $0.5$. The solar angle $\theta_{12}$ (the best measured) is large, $\sin^2{\theta_{12}}\sim 0.3$, but certainly not maximal (by more than 5$\sigma$). The third angle $\theta_{13}$, strongly limited mainly by the CHOOZ experiment, has at present a $3\sigma$ upper limit given by about $\sin^2{\theta_{13}}\leq 0.04$. The non conservation of the three separate lepton numbers and the large leptonic mixing angles make it possible that processes like $\mu \rightarrow e \gamma$ or $\tau \rightarrow \mu \gamma$ might be observable, not in the SM but in extensions of it like the MSSM. Thus, for example, the outcome of the now running experiment MEG at PSI \cite{meg}, aiming at improving the limit on $\mu \rightarrow e \gamma$ by 1 or 2 orders of magnitude, is of great interest. 

\section{Problems of the Standard Model}

No signals of new physics were
found neither in electroweak precision tests nor in flavour physics. Given the success of the SM why are we not satisfied with that theory? Why not just find the Higgs particle,
for completeness, and declare that particle physics is closed? The reason is that there are
both conceptual problems and phenomenological indications for physics beyond the SM. On the conceptual side the most
obvious problems are the proliferation of parameters, the puzzles of family replication and of flavour hierarchies, the fact that quantum gravity is not included in the SM and the related hierarchy problem. Some of these problems could be postponed to the more fundamental theory at the Planck mass. For example, the explanation of the three generations of fermions and the understanding of fermion masses and mixing angles can be postponed. But other problems, like the hierarchy problem, must find their solution in the low energy theory. Among the main
phenomenological hints for new physics we can list the quest for Grand Unification and coupling constant merging, dark matter, neutrino masses (explained in terms of L non conservation), 
baryogenesis and the cosmological vacuum energy (a gigantic naturalness problem).

\subsection{Dark matter and dark energy}

We know by now\cite{WMAP} that  the  Universe is flat and most of it is not made up of known forms of matter: while $\Omega_{tot} \sim 1$ and $\Omega_{matter} \sim 0.3$, the normal baryonic matter is only $\Omega_{baryonic} \sim 0.044$, where $\Omega$ is the ratio of the density to the critical density. Most of the energy in the Universe is Dark Matter (DM) and Dark Energy (DE) with $\Omega_{\Lambda} \sim 0.7$. We also know that most of DM must be cold (non relativistic at freeze-out) and that significant fractions of hot DM are excluded. Neutrinos are hot DM (because they are ultrarelativistic at freeze-out) and indeed are not much cosmo-relevant: $\Omega_{\nu} \lappeq 0.015$. The identification of DM is a task of enormous importance for both particle physics and cosmology. The LHC has good chances to solve this problem in that it is sensitive to a large variety of WIMP's (Weekly Interacting Massive Particles). WIMP's with masses in the 10 GeV-1TeV range with typical EW cross-sections turn out to contribute terms of $o(1)$ to $\Omega$. This is a formidable hint in favour of WIMP's as DM candidates. By comparison, axions are also DM candidates but their mass and couplings must be tuned for this purpose. If really some sort of WIMP's are a main component of DM they could be discovered at the LHC and this will be a great service of particle physics to cosmology.
Also, we have seen that vacuum energy accounts
for about 2/3 of the critical density: $\Omega_{\Lambda}\sim 0.7$\cite{tu}. Translated into familiar units this means for the energy
density $\rho_{\Lambda}\sim (2~10^{-3}~eV)^4$ or $(0.1~{\rm mm})^{-4}$. It is really interesting (and not at all understood)
that $\rho_{\Lambda}^{1/4}\sim \Lambda_{EW}^2/M_{Pl}$ (close to the range of neutrino masses). It is well known that in
field theory we expect $\rho_{\Lambda}\sim \Lambda_{cutoff}^4$. If the cut off is set at $M_{Pl}$ or even at $0(1~{\rm TeV})$
there would be an enormous mismatch. In exact SUSY $\rho_{\Lambda}=0$, but SUSY is broken and in presence of breaking 
$\rho_{\Lambda}^{1/4}$ is in general not smaller than the typical SUSY multiplet splitting. Another closely related
problem is "why now?": the time evolution of the matter or radiation density is quite rapid, while the density for a
cosmological constant term would be flat in time. If so, then how comes that precisely now the two density sources are
comparable? This suggests that the vacuum energy is not a cosmological constant term, but rather the vacuum expectation
value of some field (quintessence) and that the "why now?" problem is solved by some dynamical coupling of the quintessence field with gauge singlet fields (perhaps RH neutrinos)\cite{qui}. 

\subsection{The hierarchy problem}

The computed evolution with energy
of the effective gauge couplings clearly points towards the unification of the EW and strong forces (Grand
Unified Theories: GUT's) at scales of energy
$M_{GUT}\sim  10^{15}-10^{16}$ GeV which are close to the scale of quantum gravity, $M_{Pl}\sim 10^{19}$ GeV.  GUT's are so attractive that are by now part of our culture: they provide coupling  unification, an explanation of the quantum numbers in each generation of fermions (e.g. one generation exactly fills the 16 dimensional representation of $SO(10)$), transformation of quarks into leptons and proton decay etc. One step further and one is led to
imagine  a unified theory of all interactions also including gravity (at present superstrings provide the best attempt at such
a theory). Thus GUT's and the realm of quantum gravity set a very distant energy horizon that modern particle theory cannot
ignore. Can the SM without new physics be valid up to such large energies?  The answer is presumably not: the structure of the
SM could not naturally explain the relative smallness of the weak scale of mass, set by the Higgs mechanism at $\mu\sim
1/\sqrt{G_F}\sim  250~ GeV$  with $G_F$ being the Fermi coupling constant, with respect to $M_{GUT}$ or $M_{Pl}$. This so-called hierarchy problem is due to the instability of the SM with respect to quantum corrections. This is related to
the
presence of fundamental scalar fields in the theory with quadratic mass divergences and no protective extra symmetry at
$\mu=0$. For fermion masses, first, the divergences are logarithmic and, second, they are forbidden by the $SU(2)\bigotimes
U(1)$ gauge symmetry plus the fact that at
$m=0$ an additional symmetry, i.e. chiral  symmetry, is restored. Here, when talking of divergences, we are not
worried of actual infinities. The theory is renormalizable and finite once the dependence on the cut off $\Lambda$ is
absorbed in a redefinition of masses and couplings. Rather the hierarchy problem is one of naturalness. We can look at the
cut off as a parameterization of our ignorance on the new physics that will modify the theory at large energy
scales. Then it is relevant to look at the dependence of physical quantities on the cut off and to demand that no
unexplained enormously accurate cancellations arise. 

In the past in many cases ÒnaturalnessÓ has been a good guide in particle physics. For example, without charm and the GIM mechanism the short distance contribution to the $(K-\bar K)$ mass difference would be of order $G_F^2f_K^2m_W^2m_K$, while the correct result is of order $G_F^2f_K^2m_c^2m_K$ and, without GIM, an unnatural cancellation between long and short distance contributions would be needed. Also note that $\Lambda_{QCD} << M_{GUT}$ is natural because, due to the logarithmic running of $\alpha_s$, Òdimensional transmutationÓ brings in exponential suppression.

The hierarchy problem can be put in less abstract terms (the "little hierarchy problem"): loop corrections to the higgs mass squared are
quadratic in the cut off $\Lambda$. The most pressing problem is from the top loop.
 With $m_H^2=m^2_{bare}+\delta m_H^2$ the top loop gives 
 \begin{eqnarray}
\delta m_{H|top}^2\sim -\frac{3G_F}{2\sqrt{2} \pi^2} m_t^2 \Lambda^2\sim -(0.2\Lambda)^2 \label{top}
\end{eqnarray}
If we demand that the correction does not exceed the light Higgs mass indicated by the precision tests, $\Lambda$ must be
close, $\Lambda\sim o(1~{\rm TeV})$. Similar constraints arise from the quadratic $\Lambda$ dependence of loops with gauge bosons and
scalars, which, however, lead to less pressing bounds. So the hierarchy problem demands new physics to be very close (in
particular the mechanism that quenches the top loop). Actually, this new physics must be rather special, because it must be
very close, yet its effects are not clearly visible in the EW precision tests (the "LEP Paradox"\cite{BS}) now also accompanied by a similar "flavour paradox"\cite{isid} arising from the recent precise experimental results in $B$ and $K$ decays . The main avenues open for new physics are discussed in the following sections\cite{revnp}.

\section{Supersymmetry: the standard way beyond the SM}

Models based on supersymmetry (SUSY)\cite{Martin} are the most developed and widely known. In the limit of exact boson-fermion symmetry the quadratic divergences of bosons cancel, so that
only logarithmic divergences remain. However, exact SUSY is clearly unrealistic. For approximate SUSY (with soft breaking terms),
which is the basis for all practical models, $\Lambda$ in eq.(\ref{top}) is essentially replaced by the splitting of SUSY multiplets. In particular, the top loop is quenched by partial cancellation with s-top exchange, so the s-top cannot be too heavy. 

The Minimal SUSY Model (MSSM) is the extension of the SM with minimal particle content. To each ordinary particle a s-particle is associated with 1/2 spin difference: to each helicity state of a spin 1/2 fermion of the SM a scalar is associated (for example, the electron states $e_L$ and $e_R$ correspond to 2 scalar s-electron states). Similarly to each ordinary boson a s-fermion is associated: for example to each gluon a gluino (a Majorana spin 1/2 fermion) is related. Why not even one s-particle was seen so far? A clue: observed particles are those whose mass is forbidden by $SU(2) \bigotimes U(1)$. When SUSY is broken but $SU(2) \bigotimes U(1)$ is unbroken s-particles get a mass but particles remain massless. Thus if SUSY breaking is large we understand that no s-particles have been observed yet. It is an important fact that two Higgs doublets, $H_u$ and $H_d$, are needed in the MSSM with their corresponding spin 1/2 s-partners, to give mass to the up-type and to the down-type fermions, respectively. This duplication is needed for cancellation of the chiral anomaly and also because the SUSY rules forbid that $H_d=H^\dagger_u$ as is the case in the the SM. The ratio of their two vacuum expectation values $\tan{\beta}=v_u/v_d$ (with the SM vev $v$ being given by $v=\sqrt{v_u^2+v_d^2}$) plays an important role for phenomenology.

The most general MSSM symmetric renormalizable lagrangian would contain terms that violate baryon $B$ and lepton $L$ number conservation (which in the SM, without $\nu_R$, are preserved at the renormalizable level, so that they are "accidental" symmetries). To eliminate those terms it is sufficient to invoke a discrete parity, $R$-parity, whose origin is assumed to be at a more fundamental level, which is $+1$ for ordinary particles and $-1$ for s-partners.
The consequences of $R$-parity are that s-particles are produced in pairs at colliders, 
the lightest s-particle is absolutely stable (it is called the Lightest SUSY Particle, LSP, and is a good candidate for dark matter) and s-particles decay into a final state with an odd number of s-particles (and, ultimately, in the decay chain there will be the LSP).

The necessary SUSY breaking, whose origin is not clear, can be phenomenologically introduced through soft terms (i. e. with operator dimension $< 4$) that do not spoil the good convergence properties of the theory (renormalizability and non renormalization theorems 
are maintained). We denote by $m_{soft}$ the mass scale of the soft SUSY breaking terms. The most general soft terms compatible with the SM gauge symmetry and with $R$-parity conservation introduce more than one hundred new parameters. In general new sources of flavour changing neutral currents (FCNC) and of CP violation are introduced e.g. from s-quark mass matrices. Universality (proportionality of the mass matrix to the identity matrix for each charge sector) and/or alignment (near diagonal mass matrices) must be assumed at a large scale, but renormalization group running can still produce large effects. The MSSM does provide a viable flavour framework in the assumption of $R$-parity conservation, universality of soft masses and proportionality of trilinear soft terms to the SM Yukawas (still broken by renormalization group running). As already mentioned, observable effects in the lepton sector are still possible (e.g. $\mu \rightarrow e \gamma$ or $\tau \rightarrow \mu \gamma$). This is made even more plausible by large neutrino mixings.

How can SUSY breaking be generated? Conventional spontaneous symmetry breaking cannot occur within the MSSM and also
in simple extensions of it. Probably the soft terms of the MSSM arise indirectly or radiatively 
(loops) rather than from tree level renormalizable couplings. The prevailing idea is that it happens in a "hidden sector" through non renormalizable interactions and is communicated to the visible sector by some interactions. Gravity is a plausible candidate for the hidden sector. Many theorists consider SUSY as established at the Planck
scale $M_{Pl}$. So why not to use it also at low energy to fix the hierarchy problem, if at all possible? It is interesting that viable models exist. Suitable soft terms indeed arise from supergravity when it is spontaneoulsly broken. Supergravity is a non renormalizable SUSY theory of quantum gravity\cite{Martin}. The SUSY partner of the spin-2 graviton $g_{\mu \nu}$ is the spin-3/2 gravitino $\Psi_{i \mu}$ (i: spinor index, $\mu$: Lorentz index). The gravitino is the gauge field associated to the SUSY generator. When SUSY is broken the gravitino takes mass by absorbing the 2 goldstino components (super-Higgs mechanism). In gravity mediated SUSY breaking  typically the gravitino mass $m_{3/2}$ is of order $m_{soft}$ (the scale of mass of the soft breaking terms) and, on dimensional ground, both are given by $m_{3/2}\sim m_{soft}\sim \langle F \rangle/M_{Pl}$, where $F$ is the dimension 2 auxiliary field that takes a vacuum expectation value $\langle F \rangle$ in the hidden sector (the denominator $M_{Pl}$ arises from the gravitational coupling that transmits the breaking down to the visible sector). For $m_{soft}\sim 1~{\rm TeV}$, the scale of SUSY breaking is very large of order $\sqrt{\langle F \rangle}\sim \sqrt{m_{soft}M_{Pl}}\sim 10^{11}~ {\rm GeV}$. With ~TeV mass and gravitational coupling the gravitino is not relevant for LHC physics but perhaps for  cosmology (it could be the LSP and a dark matter candidate). In gravity mediation the neutralino is the typical LSP and an excellent dark matter candidate. A lot of missing energy is a signature for gravity mediation. 

\begin{figure}[ht]
\centerline{\epsfxsize=4.6in\epsfbox{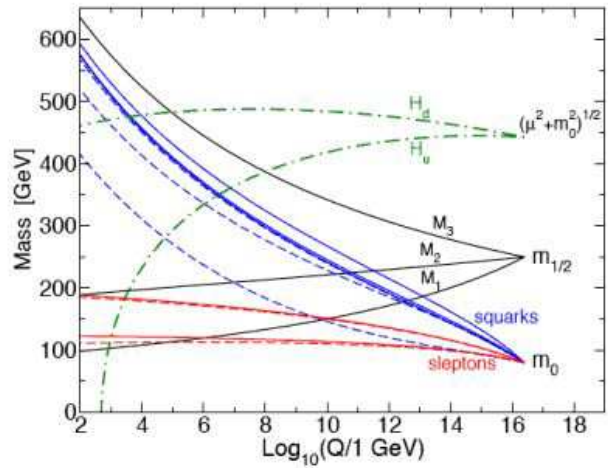}}   
\caption{A SUSY spectrum generated by universal boundary conditions at the GUT scale \label{sugra}}
\end{figure}

Different mechanisms of SUSY breaking are also being considered. In one alternative scenario\cite{gau} the (not so
much) hidden sector is connected to the visible one by ÒmessengerÓ heavy fields, with mass $M_{mess}$, which share ordinary gauge interactions and thus, in amplitudes involving only external light particles, appear in 
loops  so that $m_{soft}\sim \frac{\alpha_i}{4\pi}\frac{\langle F \rangle}{M_{mess}}$. Both gaugino and s-fermion masses are of order $m_{soft}$. Messengers can be taken in complete SU(5) representations, like 5+$\bar 5$, so that coupling unification is not spoiled. As gauge interactions are much
stronger than gravitational interactions, the SUSY breaking scale can be much smaller, as low as $\sqrt{\langle F \rangle}\sim M_{mess}\sim 10-100~{\rm TeV}$. It follows that the gravitino is very light (with mass of order or below $1~{\rm eV}$ typically) and, in these models, always is the LSP. Its couplings are observably large because the gravitino couples to SUSY particle multiplets through its spin 1/2 goldstino components. Any SUSY particle will eventually decay into the gravitino. But the decay of the next-to-the lightest SUSY particle (NLSP) could be extremely slow, with a travel path at the LHC from microscopic to astronomical distances. The main appeal of gauge mediated models is a better protection against FCNC: if one starts at $M_{mess}$ with sufficient universality/alignment then the very limited  interval for renormalization group running down to the EW scale does not spoil it. Indeed at $M_{mess}$ there is approximate alignment because the mixing
parameters $A_{u.d,l}$ in the soft breaking lagrangian are of dimension of mass and arise at two loops, so that they are suppressed.

What is unique to SUSY with respect to most other extensions of the SM is that SUSY models are well defined and computable up to $M_{Pl}$ and, moreover, are not only compatible but actually 
quantitatively supported by coupling unification and GUT's. At present the most direct
phenomenological evidence in favour of SUSY is obtained from the unification of couplings in GUT's.
Precise LEP data on $\alpha_s(m_Z)$ and $\sin^2{\theta_W}$ show that
standard one-scale GUT's fail in predicting $\alpha_s(m_Z)$ given $\sin^2{\theta_W}$ 
and $\alpha(m_Z)$ while SUSY GUT's are compatible with the present, very precise,
experimental results (of course, the ambiguities in the MSSM prediction are larger than for the SM case because of our ignorance of the SUSY spectrum). If one starts from the known values of
$\sin^2{\theta_W}$ and $\alpha(m_Z)$, one finds\cite{LP} for $\alpha_s(m_Z)$ the results:
$\alpha_s(m_Z) = 0.073\pm 0.002$ for Standard GUT's and $\alpha_s(m_Z) = 0.129\pm0.010$ for SUSY~ GUT's
to be compared with the world average experimental value $\alpha_s(m_Z) =0.118\pm0.002$\cite{pdg}. Another great asset of SUSY GUT's
is that proton decay is much slowed down with respect to the non SUSY case. First, the unification mass $M_{GUT}\sim~\rm{few}~
10^{16}~GeV$, in typical SUSY GUT's, is about 20 times larger than for ordinary GUT's. This makes p decay via gauge
boson exchange negligible and the main decay amplitude arises from dim-5 operators with higgsino exchange, leading to a
rate close but still compatible with existing bounds (see, for example,\cite{AFM}).

\begin{figure}[ht]
\centerline{\epsfxsize=4.8in\epsfbox{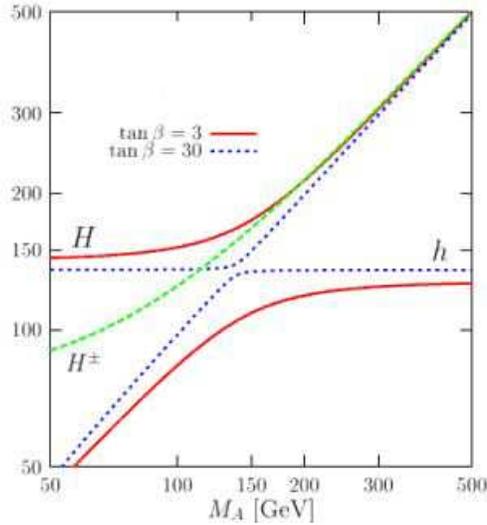}}   
\caption{The MSSM Higgs spectrum as function of $m_A$: $h$ is  the lightest Higgs, $H$ and $A$ are the heavier neutral scalar and pseudoscalar Higgs, respectively,  and $H^\pm$ are the charged Higgs bosons. The curves refer to $m_t=178~{\rm GeV}$ and large top mixing $A_t$ \label{higgs}}
\end{figure}

By imposing on the MSSM model universality constraints at $M_{GUT}$  one obtains a drastic reduction in the number of parameters at the price of more rigidity and model dependence (see Figure~\ref{sugra}\cite{Martin}). This is the SUGRA or CMSSM (C for "constrained") limit\cite{Martin}.
An interesting exercise is to repeat the fit of precision tests in the CMSSM, also including the additional data on the muon $(g-2)$, the dark matter relic density and the $b\rightarrow s \gamma$ rate. The result \cite{sus} is that the central value of the lightest Higgs mass $m_h$ goes up (in better harmony with the bound from direct searches) with moderately large $tan\beta$ and relatively light SUSY spectrum.

\begin{figure}[ht]
\centerline{\epsfxsize=5.2in\epsfbox{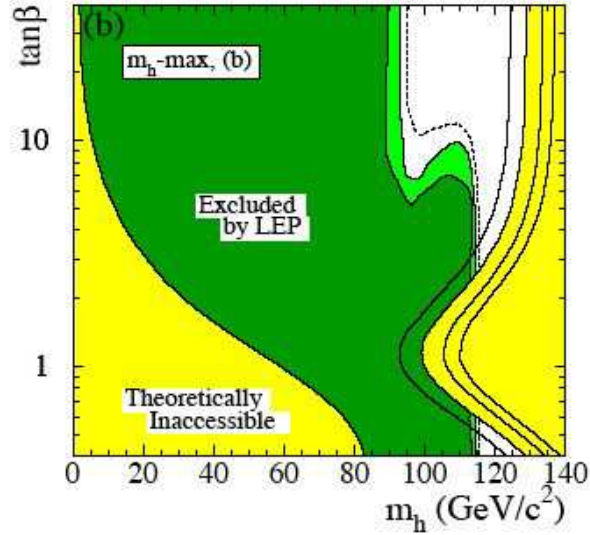}}   
\caption{Experimental limits in the $\tan{\beta}-m_h$ plane from LEP. With h one denotes the lightest MSSM Higgs boson. \label{limits}}
\end{figure}

\begin{figure}[ht]
\centerline{\epsfxsize=5.2in\epsfbox{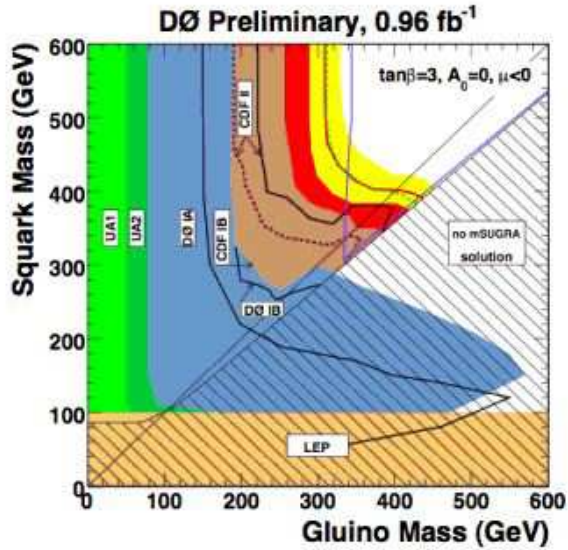}}   
\caption{Present experimental limits on s-quarks and gluinos \label{sqgl}}
\end{figure}

In spite of all these virtues it is true that the lack of SUSY signals at LEP and the lower limit on $m_H$ pose problems
for the MSSM. The predicted spectrum of Higgs particles in the MSSM is shown in Figure~\ref{higgs}\cite{djou}. As apparent from the figure the lightest Higgs particle is predicted in the MSSM to be below $m_h\lappeq~130~GeV$ (with the esperimental value of $m_t$ going down the upper limit is slightly decreased). In fact, at tree level $m^2_h=m^2_Z\cos^2{2\beta}$ and it is only through radiative corrections that $m_h$ can increase beyond $m_Z$:
\begin{eqnarray}
\Delta m_h^2\sim \frac{3G_F}{\sqrt{2} \pi^2} m_t^4\log{\frac{m_{\tilde{t_1}}m_{\tilde{t_2}}}{m^2_t}} \label{stop}
\end{eqnarray}
Here $\tilde{t}_{1,2}$ are the s-top mass eigenstates. 
The direct limit on $m_h$ from the Higgs search at LEP, shown in Figure~\ref{limits}\cite{lephwg}, considerably restricts the available parameter space of the MSSM requiring relatively large $\tan\beta$
and heavy s-top quarks. Stringent naturality constraints also follow from imposing that the EW breaking occurs at the right energy scale: in SUSY models the breaking is induced by the running of the $H_u$ mass
starting from a common scalar mass $m_0$ at $M_{GUT}$ (see Figure~\ref{sugra}). The squared Z mass $m_Z^2$ can be expressed as a linear
combination of the SUSY parameters $m_0^2$, $m_{1/2}^2$, $A^2_t$, $\mu^2$,... with known coefficients. Barring
cancellations that need fine tuning, the SUSY parameters, hence the SUSY s-partners, cannot be too heavy. The LEP limits,
in particular the chargino lower bound $m_{\chi+}\gappeq~100~GeV$, are sufficient to eliminate an important region of the
parameter space, depending on the amount of allowed fine tuning. For example, models based on gaugino universality at the
GUT scale, like the CMSSM, need a fine tuning by at least a factor of ~20. Without gaugino
universality\cite{kane} the strongest limit remains on the gluino mass: the relation reads $m_Z^2\sim 0.7~m_{gluino}^2+\dots$ and is still
compatible with the present limit $m_{gluino}\gappeq~250-300~GeV$ from the TeVatron (see Figure~\ref{sqgl}\cite{del})

This is the case of the MSSM with minimal particle content. Of course, minimality is only a simplicity assumption that could possibly be relaxed. For example, adding an additional Higgs singlet S considerably helps in addressing naturalness constraints (Next-to Minimal SUSY SM: NMSSM)\cite{nmssm},\cite{barbie}. An additional singlet can also help solving the "$\mu$- problem"\cite{Martin} . In the exact SUSY and gauge symmetric limit there is a single parameter with dimension of mass in the superpotential. The $\mu$ term in the superpotential is of the form $W_{\mu term}=\mu H_uH_d$. The mass $\mu$, which contributes to the Higgs sector masses, must be of order $m_{soft}$ for phenomenological reasons. The problem is to justify this coincidence, because $\mu$ could in principle be much larger given that it already appears at the symmetric level. A possibility is to forbid the $\mu$ term by a suitable symmetry in the SUSY unbroken limit and then generate it together with the SUSY breaking terms. For example, one can introduce a discrete parity that 
forbids the $\mu$ term. Then Giudice and Masiero\cite{gium} have observed that in general, the low energy limit of supergravity, also induces a SUSY conserving $\mu$ term together with the soft SUSY breaking terms and of the same order. A different phenomenologically appealing possibility is to replace $\mu$ with the vev of a new singlet scalar field S, thus enlarging the Higgs sector as in the NMSSM.

In conclusion the main SUSY virtues are that the hierarchy problem is drastically reduced, the model agrees with the EW data,  is consistent and computable up to $M_{Pl}$, is well compatible and indeed supported by GUT's, has good dark matter candidates and, last not least, is testable at the LHC. The delicate points for SUSY are the origin of SUSY breaking and of R-parity, the $\mu$-problem, the flavour problem and the need of sizable fine tuning. 

\section{Little Higgs models}

The non discovery of SUSY at LEP has given further impulse to the quest for new ideas on physics beyond the SM. In "little Higgs" models the symmetry of the SM is extended to a suitable global group G that also contains some
gauge enlargement of $SU(2)\bigotimes U(1)$, for example $G\supset [SU(2)\bigotimes U(1)]^2\supset SU(2)\bigotimes
U(1)$. The Higgs particle is a pseudo-Goldstone boson of G that only takes mass at 2-loop level, because two distinct
symmetries must be simultaneously broken for it to take mass,  which requires the action of two different couplings in
the same diagram. Then in the relation eq.(\ref{top})
between
$\delta m_h^2$ and
$\Lambda^2$ there is an additional coupling and an additional loop factor that allow for a bigger separation between the Higgs
mass and the cut-off. Typically, in these models one has one or more Higgs doublets at $m_h\sim~0.2~{\rm TeV}$, and a cut-off at
$\Lambda\sim~10~{\rm TeV}$. The top loop quadratic cut-off dependence is partially canceled, in a natural way guaranteed by the
symmetries of the model, by a new coloured, charge 2/3, vectorlike quark $\chi$ of mass around $1~{\rm TeV}$ (a fermion not a scalar
like the s-top of SUSY models). Certainly these models involve a remarkable level of group theoretic virtuosity. However, in
the simplest versions one is faced with problems with precision tests of the SM\cite{prob}. These problems can be fixed by complicating the model\cite{Ch}: one can introduce a parity symmetry, T-parity, and additional "mirror" fermions.  T-parity interchanges the two $SU(2)\bigotimes
U(1)$ groups: standard gauge bosons are T even while heavy ones are T odd. As a consequence no tree level contributions from heavy $WÕ$ and $ZÕ$ appear in processes with external SM particles. 
Therefore all corrections to EW observables only arise at loop level. A good feature of T-parity is that, like for R-parity in the MSSM, the lightest T-odd particle is stable (usually a B') and can be a candidate for Dark Matter (missing energy would here too be a signal) and T-odd particles are produced in pairs (unless T-parity is not broken by anomalies\cite{hill}). Thus the model could work but, in my opinion, the real limit of
this approach is that it only offers a postponement of the main problem by a few TeV, paid by a complete loss of
predictivity at higher energies. In particular all connections to GUT's are lost. Still it is very useful to offer to experiment a different example of possible new physics.

\section{Extra dimensions}

Extra dimensions  models are  among the most interesting new directions in model building. Early formulations were based on "large" extra dimensions \cite{led},\cite{Jo}. These are models with factorized metric: $ds^2=\eta_{\mu
\nu}dx^{\mu}dx^{\nu}+h_{ij}(y)dy^idy^j$, where $y^{i,j}$ denote the extra dimension coordinates and indices. Large
extra dimension models propose to solve the hierarchy problem by bringing gravity down from $M_{Pl}$ to $m\sim~o(1~{\rm TeV})$ where
$m$ is the string scale. Inspired by string theory one assumes that some compactified extra dimensions are sufficiently large
and that the SM fields are confined to a 4-dimensional brane immersed in a d-dimensional bulk while gravity, which feels the
whole geometry, propagates in the bulk. We know that the Planck mass is large just because gravity is weak: in fact $G_N\sim
1/M_{Pl}^2$, where
$G_N$ is Newton constant. The new idea is that gravity appears so weak because a lot of lines of force escape in extra
dimensions. Assume you have $n=d-4$ extra dimensions with compactification radius $R$. For large distances, $r>>R$, the
ordinary Newton law applies for gravity: in natural units, the force between two units of mass is $F\sim G_N/r^2\sim 1/(M_{Pl}^2r^2)$. At short distances,
$r\lappeq R$, the flow of lines of force in extra dimensions modifies Gauss law and $F^{-1}\sim m^2(mr)^{d-4}r^2$. By
matching the two formulas at $r=R$ one obtains $(M_{Pl}/m)^2=(Rm)^{d-4}$. For $m\sim~1~TeV$ and $n=d-4$ one finds that
$n=1$ is excluded ($R\sim 10^{15} {\rm cm}$), for $n=2~R$  is very marginal and also at the edge of present bounds $R\sim~1~ {\rm mm}$ on departures from Newton law\cite{cav}, while for $n=4,6$,
$R\sim~10^{-9}, 10^{-12}~{\rm cm}$ and these cases are not excluded.   

A generic feature of extra dimensional models is the occurrence of Kaluza-Klein (KK) modes.
Compactified dimensions with periodic boundary conditions, like the case of quantization in a box, imply a discrete spectrum with
momentum $p=n/R$ and mass squared $m^2=n^2/R^2$. In any case there are the towers of KK recurrences of the graviton. They
are gravitationally coupled but there are a lot of them that sizably couple, so that the net result is a modification
of cross-sections and the presence of missing energy. 
There are many versions of these models. The SM brane can itself have a
thickness $r$ with $r<\sim 10^{-17}{\rm cm}$ or $1/r>\sim 1{\rm TeV}$, because we know that quarks and leptons are pointlike down to
these distances, while for gravity in the bulk there is no experimental counter-evidence down to $R<\sim 0.1 {\rm mm}$ or
$1/R>\sim 10^{-3}~eV$. In case of a thickness for the SM brane there would be KK recurrences for SM fields, like $W_n$,
$Z_n$ and so on in the TeV region and above. Large extra dimensions provide an exciting scenario. Already it is remarkable that this possibility is
compatible with experiment. However, there are a number of criticisms that can be brought up. First, the hierarchy problem is
more translated in new terms rather than solved. In fact the basic relation $Rm=(M_{Pl}/m)^{2/n}$ shows that $Rm$, which one
would apriori expect to be $0(1)$, is instead ad hoc related to the large ratio $M_{Pl}/m$. Also it is
not clear how extra dimensions can by themselves solve the LEP paradox (the large top loop corrections should be
controlled by the opening of the new dimensions and the onset of gravity): since
$m_H$ is light
$\Lambda\sim 1/R$ must be relatively close. But precision tests put very strong limits on $\Lambda$. In fact in typical
models of this class there is no mechanism to sufficiently quench the corrections.

More recently models based on the Randall-Sundrum (RS) solution for the metric have attracted most of the model builders attention\cite{RS,FeAa}.  In these models the metric is not factorized and an exponential "warp" factor multiplies the ordinary 4-dimensional coordinates in the metric:
$ds^2=e^{-2kR\phi} \eta_{\mu \nu}dx^{\mu}dx^{\nu}-R^2\phi^2$  where $\phi$ is the extra coordinate. This non-factorizable metric is a solution of Einstein equations with specified 5-dimensional cosmological term. Two 4-dimensional branes are often localized at $\phi=0$ (the Planck or ultraviolet brane) and at $\phi=\pi$ (the infrared brane). In the simplest models all SM fields are located on the infrared brane. All 4-dim masses $m_4$ are scaled down with respect to
5-dimensional masses $m_5 \sim k \sim M_{Pl}$ by the warp factor: $m_4=M_{Pl}e^{-kR\pi}$. In other words mass and energies on the infrared brane are redshifted by the $\sqrt{g_{00}}$ factor. The hierarchy suppression $m_W/M_{Pl}$ could arise from the warping exponential $e^{-kR\phi}$, for not too large values of the warp factor exponent: $kR\sim 12$ (extra dimension are not "large" in this case). The question of whether these values of $kR$ can be stabilized has been discussed in ref.\cite{GW}. It was shown that the determination of $kR$ at a compatible value can be assured by a scalar field in the bulk ("radion") with a suitable potential which offer the best support to the solution of the hierarchy problem in this context. In the original RS models where the SM fields are on the brane and gravity is in the bulk there is a tower of spin-2 KK graviton resonances. Their couplings to ordinary particles are of EW order (because their propagator masses are red shifted on the infrared brane) and universal for all particles. These resonances could be visible at the LHC. Their signature is spin-2 angular distributions and universality of couplings. The RS original formulation is very elegant but 
when going to a realistic formulation it has problems, for example with EW precision tests. Also,
In a description of physics from $m_W$ to $M_{Pl}$ there should be
place for GUTÕs. But, if all SM particles are on the {\rm TeV} brane the effective theory cut-off is low and no way to $M_{GUT}$ is open. Inspired by RS different realizations of warped geometry were tried: gauge fields in the bulk and/or all SM fields (except the Higgs) on the bulk. The hierarchy of fermion masses can be seen as the result of the different profiles of the corresponding distributions in the bulk: the heaviest fermions are those closest to the brane where the Higgs is located. 
While no simple, realistic model has yet emerged as a benchmark, it is attractive to imagine that ED could be a part of the truth, perhaps coupled with some additional symmetry or even SUSY.

Extra dimensions offer new possibilities for SUSY breaking. In fact, ED can realize a geometric separation between the hidden (on the Planck brane) and the visible sector (on the TeV brane), with gravity mediation in the bulk. In anomaly mediated SUSY breaking\cite{ano} 5-dim quantum gravity effects act as messengers. The name comes because
$L_{soft}$ can be understood in terms of the anomalous violation of a local superconformal invariance. In a particular formulation of 5 dimensional supergravity, at the 
classical level, the soft term are exponentially suppressed on the MSSM
brane. SUSY breaking effects only arise at quantum level through beta
functions and anomalous dimensions of the brane couplings and fields. In this case gaugino masses are proportional to gauge coupling beta functions, so that the gluino is much heavier than the electroweak gauginos. 

In the  general context of extra dimensions an interesting direction of development is the study of symmetry breaking by orbifolding and/or boundary conditions. Orbifolding means that we have a 5 (or more) dimensional theory where the extra dimension $x_5=y$ is compactified. Along $y$ one or more $Z_2$ reflections are defined, for example $P= y \leftrightarrow -y$ 
(a reflection around the horizontal diameter) and $P'= y \leftrightarrow -y-\pi R$ (a reflection around the vertical diameter). A field $\phi(x_\mu, y)$ with definite $P$ and $P'$ parities can be Fourier expanded along $y$. Then $\phi_{++}, \phi_{+-}, \phi_{-+}, \phi_{--}$ have the n-th Fourier components proportional to $\cos{\frac{2ny}{R}}, \cos{\frac{(2n+1)y}{R}}, \sin{\frac{(2n+1)y}{R}}, \sin{\frac{(2n+2)y}{R}} $, respectively. On the branes located at the fixed points of $P$ and $P'$, $y=0$ and $y= -\pi R/2$, the symmetry is reduced: indeed at $y=0$ only $\phi_{++}$ and $\phi_{+-}$ are non vanishing and only $\phi_{++}$ is massless. 

For example, at the GUT scale, symmetry breaking by orbifolding can be applied to obtain a reformulation of SUSY GUT's where many problematic features of ordinary GUT's (e.g. a baroque Higgs sector, the doublet-triplet splitting problem, fast proton decay etc) are eliminated or improved\cite{Kaw},\cite{edgut}. In these GUT models the metric is factorized, but while for the hierarchy problem $R\sim 1/{\rm TeV}$, here one considers $R\sim 1/M_{GUT}$ (not so large!). $P$ breaks $N=2$ SUSY, valid in 5 dimensions, down to $N=1$ while $P'$ breaks SU(5). At the weak scale there are models where SUSY, valid in $n>4$ dimensions, is broken by orbifolding\cite{ant}, in particular the model of ref.\cite{bar}, where the mass of the Higgs is in principle computable and is predicted to be light.

Symmetry breaking by boundary conditions (BC) is more general than the particular case of orbifolding\cite{groj}. Breaking by orbifolding is somewhat rigid: for example, normally the rank remains fixed and it corresponds to Higgs bosons in the adjoint representation (the role of the Higgs is taken by the 5th component of a gauge boson). BC allow a more general breaking pattern and, in particular, can lower the rank of the group. In a simplest version one starts from a 5 dimensional model with two branes at $y=0,~\pi R$. In the action there are terms localised on the  branes that also should be considered in the minimization procedure. For a scalar field $\varphi$ with a mass  term ($M$) on the boundary, one obtains the  Neumann BC $\partial_y \varphi=0$ for $M\rightarrow 0$ and the Dirichlet BC $\varphi=0$ for $M\rightarrow \infty$. In gauge theories one can introduce Higgs fields on the brane that take a vev. The crucial property is that the gauge fields take a mass as a consequence of the Higgs mechanism on the boundary but the mass remains finite when the Higgs vev goes to infinity. Thus the Higgs on the boundary only enters as a way to describe and construct the breaking but actually can be removed and still the gauge bosons associated to the broken generators take a finite mass. One is then led to try to formulate "Higgsless models" for EW symmetry breaking based on BC\cite{Hless}. The RS warped geometry can be adopted with the Planck and the infrared branes. There is a larger gauge symmetry in the bulk which is broken down to different subgroups on the two branes so that finally of the EW symmetry only $U(1)_Q$ remains unbroken. The $W$ and $Z$ take a mass proportional to $1/R$. Dirac fermions are on the bulk and only one chirality has a zero mode on the SM brane. In Higgsless models unitarity, which in general is violated in the absence of a Higgs, is restored by exchange of infinite KK recurrences, or the breaking is delayed by a finite number, with cancellations guaranteed by sum rules implied by the 5-dim symmetry. Actually no compelling, realistic Higgsless model for EW symmetry breaking emerged so far. There are serious problems from EW precision tests \cite{BPR} because the smallness of the $W$ and $Z$ masses forces $R$ to be rather small and, as a consequence, the spectrum of KK recurrences is quite close. However these models are interesting as rare examples where no Higgs would be found at the LHC but instead new signals appear (new vector bosons, i.e. KK recurrences of the $W$ and $Z$).

An interesting model that combines the idea of the Higgs as a Goldstone boson and warped extra dimensions was proposed and studied in references\cite{con} with a sort of composite Higgs in a 5-dim AdS theory. It can be considered as a new way to look at walking technicolor\cite{L-C} using AdS/CFT correspondence. In a RS warped metric framework all SM fields are in the bulk but the Higgs is localised near the TeV brane. The Higgs is a pseudo-Goldstone boson (as in Little Higgs models) and EW symmetry breaking is triggered by 
top-loop effects. In 4-dim the bulk appears as a strong sector.  The 5-dimensional theory is weakly coupled so that the Higgs potential and EW observables can be computed.
The Higgs is rather light: $m_H < 185~{\rm GeV}$. Problems with EW precision tests and the $Zb \bar b$ vertex have been fixed in latest versions. The signals at the LHC for this model are 
a light Higgs and new resonances at ~1- 2 TeV

In conclusion, note that apart from Higgsless models (if any?) all theories discussed 
here have a Higgs in LHC range (most of them light).

\section{Effective theories for compositeness}  

In this approach\cite{comp} a low energy theory from truncation of some UV completion is described in terms of an elementary sector (the SM particles minus the Higgs), a composite sector (including the Higgs, massive vector bosons $\rho_\mu$ and new fermions) and a mixing sector. The Higgs is a pseudo Goldstone boson of a larger broken gauge group, with $\rho_\mu$ the corresponding massive vector bosons. Mass eigenstates are mixtures of elementary and composite states, with light particles mostly elementary and heavy particles mostly composite. But the Higgs is totally composite (perhaps also the right-handed top quark). New physics in the composite sector is well hidden because light particles have small mixing angles. The Higgs is light because only acquires
mass through interactions with the light particles from their composite components. This general description can apply to models with a strongly interacting sector as arising from little Higgs or extra dimension scenarios.

\section{The anthropic solution}

The apparent value of the cosmological constant $\Lambda$ poses a tremendous, unsolved naturalness problem\cite{tu}. Yet the value of $\Lambda$ is close to the Weinberg upper bound for galaxy formation\cite{We}. Possibly our Universe is just one of infinitely many (Multiverse) continuously created from the vacuum by quantum fluctuations. Different physics takes place in different Universes according to the multitude of string theory solutions (~$10^{500}$). Perhaps we live in a very unlikely Universe but the only one that allows our existence\cite{anto}. I find applying the anthropic principle to the SM hierarchy problem excessive. After all we can find plenty of models that easily reduce the fine tuning from $10^{14}$ to $10^2$: why make our Universe so terribly unlikely? By comparison the case of the cosmological constant is a lot different: the context is not as fully specified as the for the SM (quantum gravity, string cosmology, branes in extra dimensions, wormholes through different Universes....)

\section{Conclusion}

Supersymmetry remains the standard way beyond the SM. What is unique to SUSY, beyond leading to a set of consistent and
completely formulated models, as, for example, the MSSM, is that this theory can potentially work up to the GUT energy scale.
In this respect it is the most ambitious model because it describes a computable framework that could be valid all the way
up to the vicinity of the Planck mass. The SUSY models are perfectly compatible with GUT's and are actually quantitatively
supported by coupling unification and also by what we have recently learned on neutrino masses. All other main ideas for going
beyond the SM do not share this synthesis with GUT's. The SUSY way is testable, for example at the LHC, and the issue
of its validity will be decided by experiment. It is true that we could have expected the first signals of SUSY already at
LEP, based on naturality arguments applied to the most minimal models (for example, those with gaugino universality at
asymptotic scales). The absence of signals has stimulated the development of new ideas like those of extra dimensions
and "little Higgs" models. These ideas are very interesting and provide an important reference for the preparation of LHC
experiments. Models along these new ideas are not so completely formulated and studied as for SUSY and no well defined and
realistic baseline has sofar emerged. But it is well possible that they might represent at least a part of the truth and it
is very important to continue the exploration of new ways beyond the SM. New input from experiment is badly needed, so we all look forward to the start of the LHC.

\section*{Acknowledgments}

I wish to most warmly thank the Organizers of the Lake Louise Winter Institute, in particular Faqir Khanna,  Andrzej Czarnecki and Roger Moore for their kind invitation and hospitality. I am also grateful to Dr. Paolo Gambino for providing me with an update of Figs. 2 and 3.

\end{document}